\documentclass{PoS}

\def\ra{\rightarrow}
\newcommand{\GeV}{\ensuremath{\mathrm{Ge\kern -0.1em V}}}
\newcommand{\MeV}{\ensuremath{\mathrm{Me\kern -0.1em V}}}

\newcommand{\bs}{\ensuremath{B_s^0}}
\newcommand{\bd}{\ensuremath{B^0}}
\newcommand{\bsd}{\ensuremath{B_s^0(B^0)}}
\newcommand{\bu}{\ensuremath{B^{+}}}

\newcommand{\mm}{\ensuremath{\mu^{+}\mu^{-}}}

\newcommand{\bsmm}{\ensuremath{\bs\ra\mm}}
\newcommand{\bdmm}{\ensuremath{\bd\ra\mm}}
\newcommand{\bsdmm}{\ensuremath{\bsd\ra\mm}}
\newcommand{\bjk}{\ensuremath{\bu\ra J/\psi K^{+}}}

\newcommand{\bsjp}{\ensuremath{\bs\ra J/\psi \phi}}

\newcommand{\brbsmm}{\ensuremath{\mathcal{B}(\bsmm)}}
\newcommand{\brbdmm}{\ensuremath{\mathcal{B}(\bdmm)}}
\newcommand{\Mmm}{\ensuremath{m_{\mu\mu}}}

\newcommand{\ctau}{\ensuremath{\lambda}}
\newcommand{\pting}{\ensuremath{\Delta\Theta}}
\newcommand{\iso}{\ensuremath{\mathit{I}}}
\newcommand{\nn}{\ensuremath{\nu_{N}}}
\newcommand{\ptmm}{\ensuremath{\vec{p}^{\:\mu\mu}_{T}}}
\newcommand{\pmm}{\ensuremath{\vec{p}^{\:\mu\mu}}}
\newcommand{\cdf}{CDF~II}

\newcommand{\dedx}{\ensuremath{dE/dx}}

\newcommand{\Do}{D\O}
\newcommand{\bmms}{\ensuremath{b \ra\ s\ \mm}}
\newcommand{\bummkplus}{\ensuremath{\bu\ \ra\ K^{+} \mm}}
\newcommand{\bdmmkstar}{\ensuremath{\bd\ \ra\ K^{*0} \mm}}
\newcommand{\bsmmphi}{\ensuremath{\bs\ \ra\ \phi\ \mm}}
\newcommand{\lbmmlzero}{\ensuremath{\Lambda^{0}_{b} \ra\ \Lambda\ \mm}}

\newcommand{\dzero}{\ensuremath{D^0}}
\newcommand{\dmm}{\ensuremath{\dzero\ \ra \mm}}
\newcommand{\brdmm}{\ensuremath{\mathcal{B}(\dmm)}}
\newcommand{\dpipi}{\ensuremath{\dzero\ \ra \pi^{+}\pi^{-}}}
\newcommand{\dkpi}{\ensuremath{\dzero\ \ra K^{-}\pi^{+}}}

\title{Flavor Changing Neutral Current at the Tevatron}

\ShortTitle{Flavor Changing Neutral Current at the Tevatron}

\author{\speaker{Michael J. Morello}\thanks{On behalf of the CDF and \Do\ Collaborations.}\\
        Fermi National Accelerator Laboratory\\
        E-mail: \email{morello@fnal.gov}}

\abstract{Processes involving flavor changing neutral currents (FCNC) provide excellent
signatures with which to search for evidence of new physics. They have 
very small branching fractions in the Standard Model since they are highly suppressed 
by Glashow-Iliopoulos-Maiani (GIM) mechanism. They occur only through higher order diagrams, and 
new particles contributions can provide a significant enhancements, which would be 
an uniquevocal signs of physics beyond the Standard Model.
In this paper we present the most recent measurements on FCNC processes
performed by CDF and \Do\ Collaborations, while last section is devote to the charm physics at CDF.}

\FullConference{12th International Conference on B-Physics at Hadron Machines - BEAUTY 2009\\
		 September 07 - 11 2009\\
		 Heidelberg, Germany}

\begin{document}

\section{Search for  rare $ \bs (\bd) \rightarrow \mu^+\mu^-$ decay modes}

The FCNC decays $\bs (\bd) \rightarrow \mu^+\mu^-$~\cite{qconj}
occur in the Standard Model (SM) only through higher order diagrams and are
further suppressed by the helicity factor, $(m_\mu/m_B)^2$. 
The $\bd$ decay is also suppressed with respect to the $\bs$ decay
by the ratio of CKM elements, $\left|V_{td}/V_{ts}\right|^2$.
The SM expectations for these branching fractions are 
$\brbsmm = (3.42\pm0.54)\times10^{-9}$ and $\brbdmm = (1.00\pm0.14)\times10^{-10}$~\cite{smbr}, 
which are one order of magnitude smaller than current experimental sensitivity.
Enhancements to \bsdmm\ occur in many new physics models~\cite{tanb}.
In the absence of an observation, limits on \brbsmm\ are complementary to those 
provided by other experimental measurements, and together would significantly constrain 
the allowed supersymmetric parameter space. 
In general, the search for these rare decays is central to exploring a 
large class of new physics models.
In the following  we will present recent results on the search of these very rare decays 
from CDF~\cite{Aaltonen:2007kv,cdf_9892} and
\Do~\cite{d0_5344,d0_5906} Collaborations. 
Since two analyses are very similar, we will shortly describe
only CDF analysis. Details on CDF and \Do\ detectors can be found in~\cite{beauty_kreps}.

At CDF in the offline analysis, the trigger selection~\cite{Aaltonen:2007kv,cdf_9892} is refined by
applying a series of ``baseline'' requirements that substantially reduce
the backgrounds while preserving the majority of the signal.
We select two oppositely charged muon candidates within a dimuon 
invariant mass window of $4.669 < \Mmm < 5.969\:\GeV/c^2$ around
the $\bs$ and $\bd$ masses. 
Backgrounds from hadrons misidentified as muons
are suppressed by selecting muon candidates using a likelihood
function.  This function tests the consistency of electromagnetic and
hadronic energy with that expected for a minimum  ionizing particle
and the differences between extrapolated track trajectories and muon system hits~\cite{Aaltonen:2007kv}.
In addition, backgrounds from kaons that penetrate through the calorimeter to the muon system or decay in flight outside
the drift chamber are further suppressed by a loose selection based on the measurement
of the ionization per unit path length, $\dedx$~\cite{bhhprl,bhhprl2}.
The inputs to the muon likelihood and the $\dedx$ performance are calibrated using
samples of $J/\psi\ra\mm$, $D^0 \rightarrow K^- \pi^+$ and $\Lambda \rightarrow p\pi^-$ decays.
To reduce combinatorial backgrounds the muon candidates are required to have transverse momentum 
relative to the beam direction 
$p_{T}>2.0\;\GeV/c$, and  $| \ptmm | > 4\;\GeV/c$, where $\ptmm$ is the transverse component of the
sum of the muon momentum vectors.
The remaining pairs of muon tracks are fit under the constraint that they
come from the same three-dimensional (3D) space point.
To achieve further separation of signal from background, 
we employ additional discriminating variables.
These include the measured proper decay time, $\ctau$; the proper decay time divided by the estimated uncertainty, 
$\ctau / \sigma_{\ctau}$; the 3D opening angle 
between vectors \pmm\ and the displacement vector between the primary vertex and the dimuon vertex, $\pting$;
and the {\it B}-candidate track isolation, $\iso$~\cite{isodef}.  We require 
that $\ctau / \sigma_{\ctau} > 2$,
$\pting < 0.7~\rm{rad}$, and $\iso > 0.50$.
The baseline selection reduces combinatorial backgrounds 
by a factor of 300 while keeping approximately 50\% 
of the signal events that are within the acceptance (geometric and kinematic requirements) of the trigger.
A sample of $\bjk$ events is collected to serve as a normalization mode using
the same baseline
requirements, but including a requirement of $p_T>1~\GeV/c$ for the kaon  
candidate and constructing the $\bjk\ra\mm$ vertex using only the muon candidate tracks.

\begin{figure}[top]
  \centering
   \includegraphics[scale=0.31]{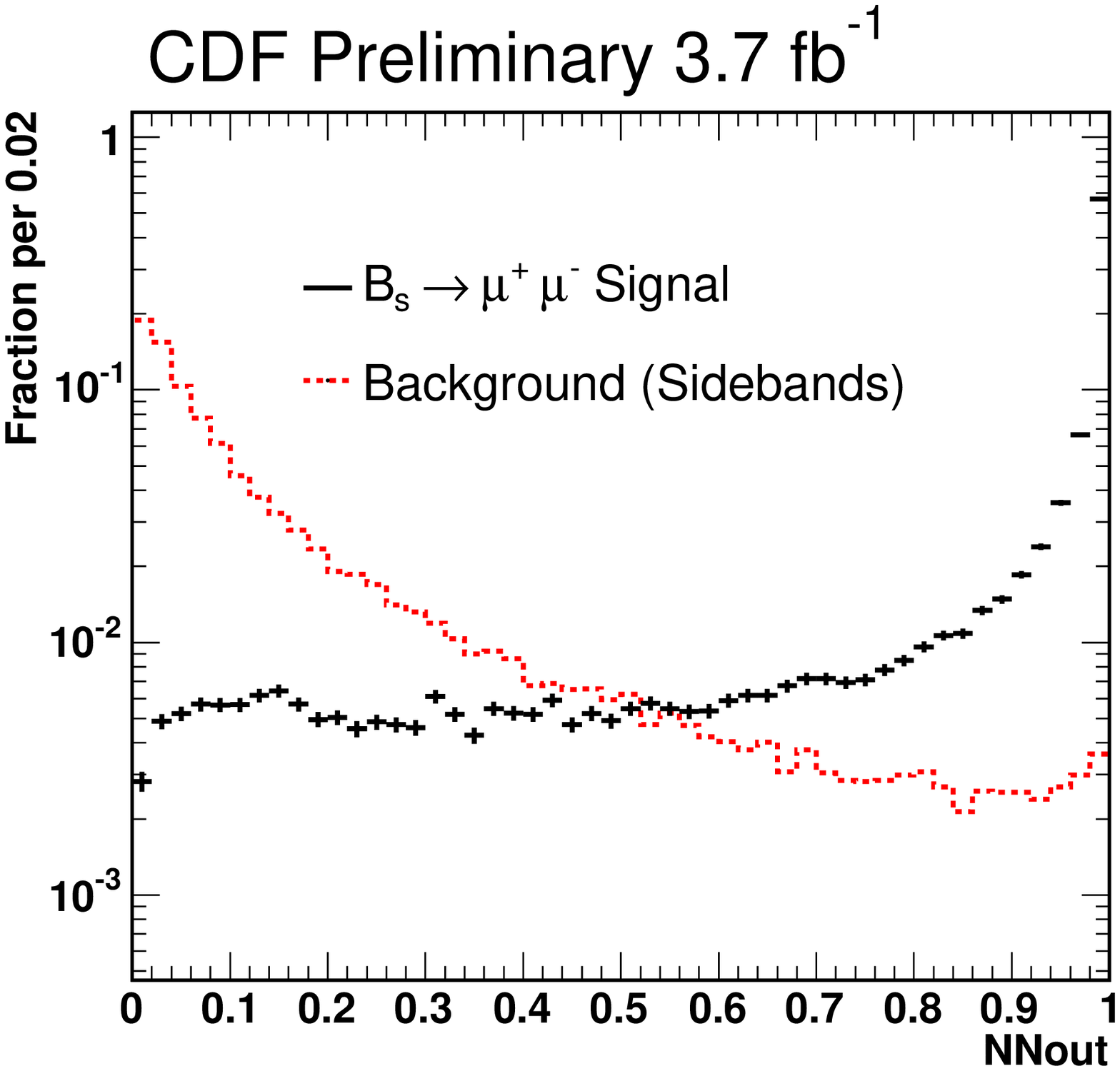}    
\hspace{0.5cm}  
   \includegraphics[scale=0.31]{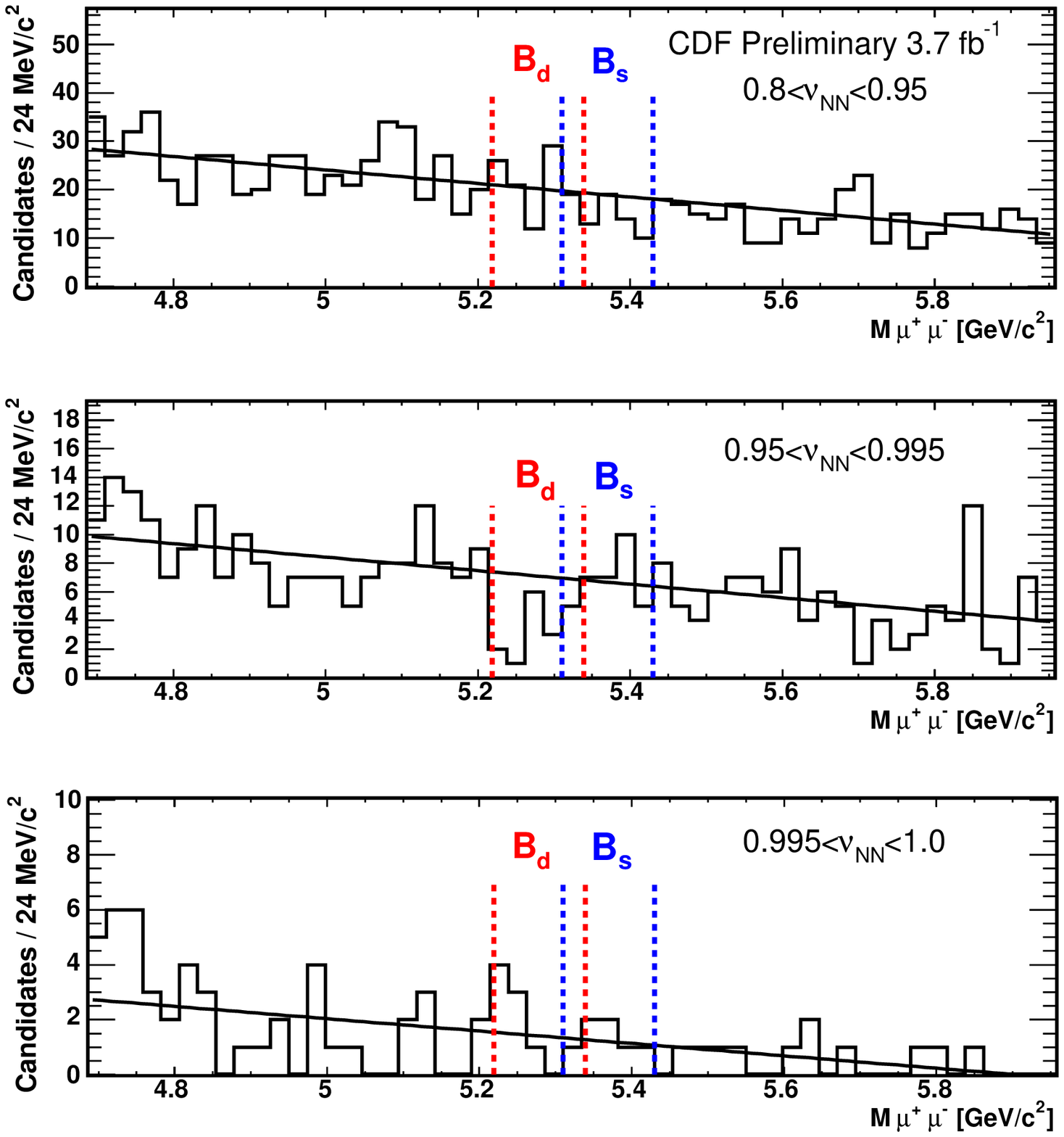}
  \caption{\label{fig:bsmm} Distributions of \nn\ for simulated \bsmm\ signal  
and observed sideband events (on the left). The \mm\ invariant mass distribution for events 
   satisfying all selection criteria for the final three ranges of $\nu_{N}$ (on the right).} 
\end{figure}

For the final event selection we use the following discriminating variables:
\Mmm , \ctau , $\ctau / \sigma_{\ctau}$, \pting , \iso , $| \ptmm |$, 
and the $p_T$ of the lower momentum muon
candidate.  
To enhance signal and background separation we construct a
NN discriminant, \nn, based on all the discriminating variables except $\Mmm$, which
is used to define signal and sideband background regions.
The NN is trained using background events sampled from the sideband regions
and signal events generated with a simulation described below.
The \nn\ distributions of $\bs$ signal and sideband background
events are shown in fig.~\ref{fig:bsmm}.  
%

For measuring efficiencies, estimating backgrounds, and optimizing the
analysis, samples of $\bsdmm$, $\bjk$, and $B \ra h^+h^-$ (where $h^{\pm}$ are $\pi^{\pm}$ or $K^{\pm}$) 
are generated with the {\sc pythia} simulation 
program~\cite{pythia} and a \cdf\ detector simulation.
The $B$-hadron $p_{T}$ 
spectrum and the \iso\ distribution of the $B$-hadrons 
are weighted to match distributions measured in samples of
\bjk\ and \bsjp\ decays. 

We use a relative normalization to determine the \bsmm\ branching
fraction:
\begin{eqnarray}\label{eq:brbsmm}
\brbsmm = \frac{N_s}{N_{+}}\cdot
     \frac{\alpha_{+}}{\alpha_{s}}
     \cdot\frac{\epsilon_{+}}{\epsilon_{s}} \cdot 
     \frac{1}{\epsilon_{N}} \cdot \nonumber
     \frac{f_{u}}{f_{s}}  
     \cdot \mathcal{B}(B^+),~~~ \rm{(1)}
\end{eqnarray}
where  $N_s$ is the number of \bsmm\ candidate events.
We observe about $N_{+} \approx 19,700$
\bjk\ candidates. 
We use $\mathcal{B}(B^+) = \mathcal{B}(\bjk\ra\mm K^{+})=(5.94\pm0.21)\times 10^{-5}$~\cite{PDG2006} and
the ratio of $B$-hadron production fractions ${f_{u}}/{f_{s}}=3.86\pm0.59$~\cite{PDG2006}. 
The parameter $\alpha_{s}$ ($\alpha_{+}$) is the acceptance of the trigger and  
$\epsilon_{s}$ ($\epsilon_{+}$) is the efficiency of the
reconstruction requirements for the signal (normalization) mode.  The reconstruction efficiency
includes trigger, track, muon, and baseline selection efficiencies.  
The NN efficiency, ${\epsilon_{N}}$, only applies to the signal mode.
The expression for \brbdmm\ is derived by replacing \bs\ with \bd\ 
and the fragmentation ratio with ${f_{u}}/{f_{d}}=1$.

The ratio of acceptances ${\alpha_{+}}/{\alpha_{s}}$ 
is measured using simulated events, and its uncertainty includes contributions from systematic variations
of the modeling of the $B$-hadron $p_{T}$ distributions, the longitudinal 
beam profile, and from the statistics of the simulated event samples. 
The ratio of reconstruction efficiencies ${\epsilon_{+}}/{\epsilon_{s}}$ 
is measured using simultaneously data and simulation. 
Muon reconstruction efficiencies are estimated as a function of muon $p_T$  
using observed event samples of inclusive $J/\psi\ra\mm$ decays.
Systematic uncertainties in this ratio 
largely cancel with the exception of the kaon efficiency from the \bu\ decay.
The uncertainty  is dominated by kinematic differences
between inclusive $J/\psi\ra\mm$ and \bsdmm\ decays.
The efficiency, $\epsilon_{N}$, is estimated from the simulation.  
We assign a relative systematic uncertainty on $\epsilon_{N}$ of 6\%\ 
based on comparisons of NN performance in simulated and observed $\bjk$ event samples and 
the statistical uncertainty on studies of the $\bs$ $p_T$ and $\iso$ distributions from observed $\bsjp$ decays.

The expected background is obtained by summing contributions from
the combinatorial continuum and from $B\rightarrow h^+h^-$ decays, 
which peak in the $\bs$ and $\bd$ invariant mass signal region and do not occur in the sidebands.
The contribution from
other heavy-flavor decays is negligible.  We estimate the combinatorial
background by linearly extrapolating from the sideband region to the
signal region.
The $B\rightarrow h^+h^-$ contributions are about a factor of ten smaller
than the combinatorial background and are estimated using efficiencies
taken from the simulation,  probabilities of misidentifying hadrons as muons measured in
a $D^0 \rightarrow \pi K$ data sample, and normalizations derived from branching 
fractions from refs.~\cite{bhhprl,PDG2006}.
The two-body invariant mass distribution of the simulated $B\rightarrow h^+h^-$ candidates is
calculated from the momentum of the hadrons assuming the muon mass hypothesis.
The background estimates are cross-checked using
three independent control samples: $\mu^{\pm}\mu^{\pm}$ events, \mm\ events with $\ctau<0$, and
a misidentified muon-enhanced \mm\ sample in which we require one muon 
candidate to fail the muon quality requirements.  
We compare the predicted and observed
number of events in these samples for a wide range of \nn\
requirements and observe no significant discrepancies.


The $\mm$ invariant mass distributions for the three different $\nu_{N}$ ranges
are shown in fig.~\ref{fig:bsmm}. The observed event rates are
consistent with SM background expectations.
Using a data sample of 3.7~fb$^{-1}$ of integrated luminosity, collected by \cdf\ experiment,
 we extract 95\% (90\%) C.L. limits of $\brbsmm < 4.3\times 10^{-8}$ $(3.6\times 10^{-8})$~\cite{cdf_9892} and
$\brbdmm < 7.6\times 10^{-9}$ $(6.0\times 10^{-9})$~\cite{cdf_9892}, which are currently the world's best upper limits
for both processes.  
Assuming $N_s=1$, from eq.~\ref{eq:brbsmm}  
we compute the single event sensitivity (SES) for the \bsmm\ decay mode as 
$3.2 \times 10^{-9}$ for all mass and $\nu_{N}$ bins. 
The SES for the  \bsmm\ is smaller than the expected SM branching 
fraction and we expect 1.2 events from \bsmm\ decays with 0.7 events occurring 
in the highest sensitivity $\nu_{N}$ bin. 
The result for \bsmm\ is slightly in excess of the expected limit, which is
$\brbsmm\ < 3.3 \times 10^{-8} (2.7 \times 10^{-8})$ at 95(90)\%~C.L., and this was estimated with
 no signal hypothesis. We calculate the P-value of the excess as 23\% corresponding to $0.73\sigma$. 


\Do\ performs a similar analysis. Latest results are based on a data sample of 
integrated luminosity of 2~fb$^{-1}$. The observed event rates are, also in this case, 
consistent with SM background expectations. We extract 95(90)\% C.L. limit of
$\brbsmm < 9.3\times 10^{-8}$ $(7.5\times 10^{-8})$~\cite{d0_5344}. 
With 5~fb$^{-1}$ of data collected by the \Do\ experiment, we have studied the sensitivity to the 
branching fraction of \bsmm\ decays. An expected upper limit on the branching fraction is 
$\brbsmm\ < 5.3(4.3) \times 10^{-8}$ at the 95(90)\% C.L.~\cite{d0_5906}.

We observe no evidence for new physics and set limits that are
the most stringent to date, improving the previous results~\cite{Aaltonen:2007kv,d0_5344,Aubert:2007hb} 
by a factor of 1.5 or more. These limits place further constraints on
new physics models,  
and complement its direct searches. We expect the analysis 
sensitivity to continue to improve as we include larger data sets. 

\section{$b \to s\ \ell^+\ell^-$  transition at the Tevatron}

\begin{figure}[top]
  \centering
   \includegraphics[scale=0.25]{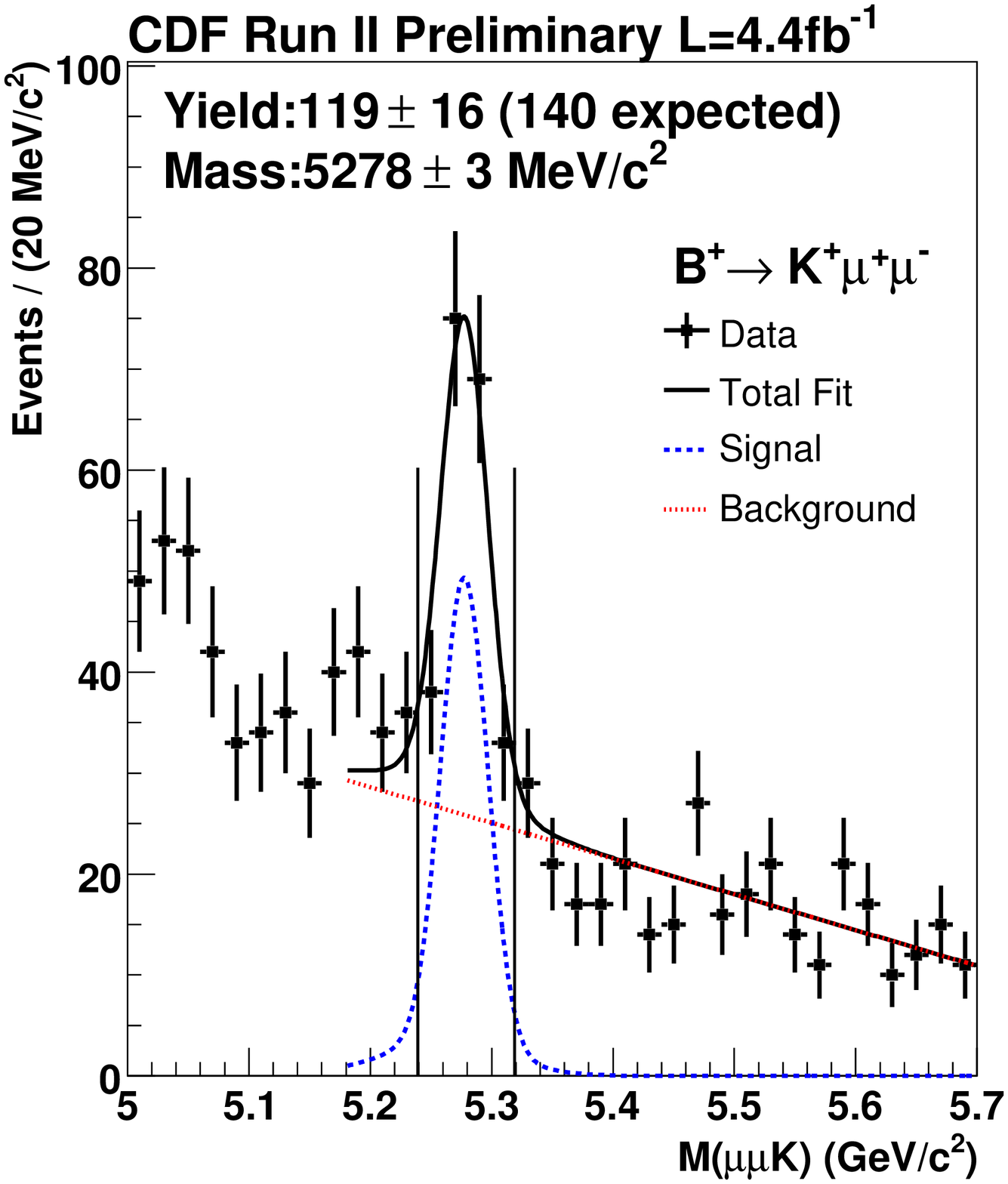}    
\hspace{0.5cm}  
   \includegraphics[scale=0.25]{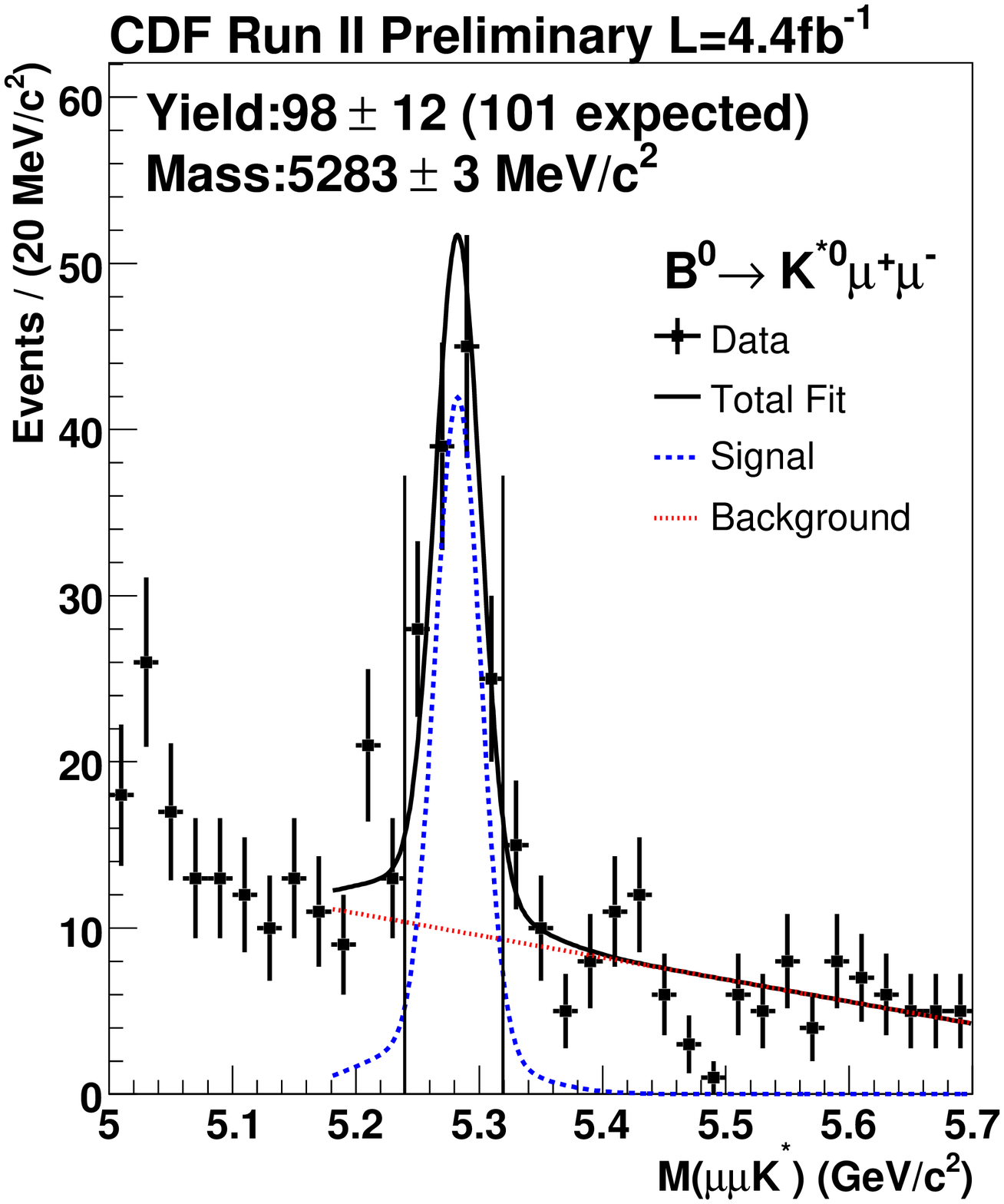}
\hspace{0.5cm}  
   \includegraphics[scale=0.25]{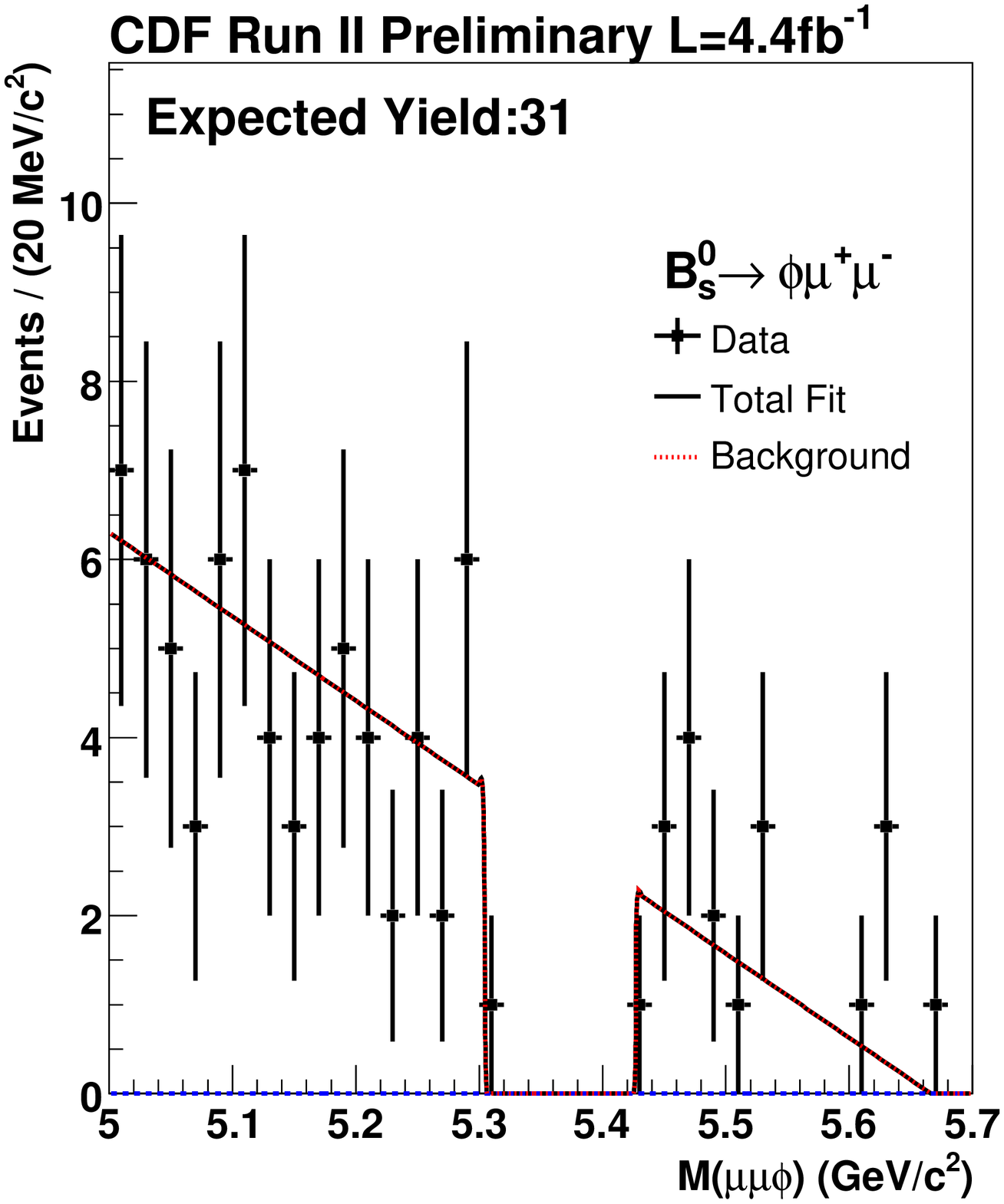}
  \caption{\label{fig:btosmumu} The $B$ invariant mass of 
\bummkplus(on the left), \bdmmkstar(center) and \bsmmphi(on the right) 
on a data sample of  4.4~fb$^{-1}$ of integrated luminosity. 
Solid, dashed, and dotted line shows total fit, signal 
PDF and background PDF, respectively, while dots with error bars are data. 
}
\end{figure}

The $b \to s\ \ell^+\ell^-$ transition is a FCNC process, which, in the SM, 
proceeds at lowest order via either a $Z$/$\gamma$ penguin diagram or a $W^+ W^-$ box diagram, as the \bsdmm\ decays. 
The effective Wilson coefficients $C_7$, $C_9$, and $C_{10}$ 
describe the amplitudes from the electromagnetic penguin, 
the vector electroweak, and the axial-vector electroweak contributions, respectively. 
These amplitudes may interfere with the contributions from non-SM particles~\cite{nonsm}, 
therefore the transition can probe the presence of yet unobserved particles and processes. 
More specifically, the lepton forward-backward asymmetry ($A_{FB}$) 
and the differential branching fraction as functions of dilepton invariant mass ($M_{\ell\ell}$) 
in the decays $B \to K^{*} \ell^+\ell^-$ differ from the SM expectations in various extended models~\cite{kll_th}.
The former is largely insensitive to the theoretical uncertainties of the form factors describing the decay, 
and can hence provide a stringent experimental test of the SM. 
The latter has been already determined~\cite{kll_exp_bl,kll_exp_bb}, however the precision is still limited 
by statistics.
It can be used to extract the information on 
the coefficients associated with the theoretical models as well.
For more details see ref.~\cite{beauty_gudrun}.

Although the branching ratio of each \bmms\ decay 
is quite small as $\mathcal{O}(10^{-7})$, the decay is experimentally 
clean due to opposite sign muons. 
Among many \bmms\ decays, the exclusive channels \bummkplus\ and \bdmmkstar\ 
have been observed and studied at Belle \cite{kll_exp_bl} and BaBar \cite{kll_exp_bb}. However, the analogous decays 
\bsmmphi\ and \lbmmlzero\ have not been observed despite searches by CDF \cite{Aaltonen:2008xf} and \Do~\cite{Abazov:2006qm}. 
Electrons can be reconstructed at the TeVatron, but they involve additional experimental difficulties 
and they will be treated in next iterations of the analysis.

CDF is currently updating the analysis on a data sample of 4.4~fb$^{-1}$ of integrated luminosity~\cite{hideki}. 
We obtain a signal yield for the $\bdmmkstar$ decay mode equal to $98 \pm 12$ (101 expected), 
with a statistical significance of about $10 \sigma$ and a 
signal yield for the $\bummkplus$ equal to $119 \pm 16$ (140 expected), with a statistical 
significance of about $9 \sigma$. Significances are deterimined from the likelihood ratio to a null signal hypothesis. 
Search of the $\bsmmphi$ is still blind as shown in fig.~\ref{fig:btosmumu}. We expect a significant signal yield, 
also in this case, of about 31 events. Measurement of $\mathcal{B}(M_{\ell\ell}^{2})$ and $A_{FB}$ is underway~\cite{hideki_hcp}.

\section{Search for  rare \dmm\ decay mode}

The FCNC decay \dmm\ is highly suppressed  
in the SM by the nearly exact Glashow-Iliopoulos-Maiani cancellation. 
SM expects the branching fraction to be about $10^{-18}$~\cite{Burdman:2001tf} from short-distance 
 processes, increasing to about $4 \times 10^{-13}$~\cite{Burdman:2001tf} 
by including long-distance processes. The prediction 
is many orders of magnitude beyond the reach of the present generation of experiments. The 
best published upper bound is $1.3 \times 10^{-6}$ at the 90\% C.L. from BaBar~\cite{Aubert:2004bs}, 
while the world's best upper bound is $1.4 \times 10^{-7}$ at the 90\% C.L. from Belle~\cite{belle_dmm}.
 
However, new physics contributions can significantly enhance the branching ratio~\cite{Burdman:2001tf}. 
Some of these new scenarios could enhance the branching fraction to 
 the range of $10^{-8}$ to $10^{-10}$ , and in particular R-parity violating SUSY could lift the branching 
 fraction up to the level of the existing experimental bound. Similar enhancements can occur  
in $K$ and $B$ decays, but charm decays provide a unique laboratory to search for new physics 
couplings in the up-quark sector. Ref.~\cite{Golowich:2009ii} shows that, in some scenarios, 
new physics contributions to $\dzero-\overline{D}^0$ mixing can dominate, but yield a result consistent with the SM, 
the same new physics can contribute to \dmm, yielding a visible signal. 

Using a data sample of 360~pb$^{-1}$ of integrated luminosity CDF searches for \dmm\ decays.
A displaced-track trigger selects long-lived \dzero\ candidates in the \mm, $\pi^{+}\pi^{-}$, and 
$K^{-}\pi^{+}$ decay modes. We use the kinematically similar \dpipi\ channel for normalization, and the 
Cabibbo-favored \dkpi\  channel to optimize the selection criteria in an unbiased manner. 
We set an upper limit on the branching fraction  $\brdmm\ < 3.0 \times 10^{-7} (2.1 \times 10^{-7})$ at the 
95(90\%) C.L. \cite{cdf_9226}. CDF is currently analyzing a data sample of 
about 5~fb$^{-1}$ of data (already on tape), and it will integrate
8(10)~fb$^{-1}$ by the end of 2010(2011). We expect to significantly approach, by the end of Run~II, 
the interesting region $10^{-8}$ for the \dmm\ decay.

\section{Charm Mixing}

In the SM, the decay $D^0 \rightarrow K^+\pi^-$ proceeds
through a doubly Cabibbo-suppressed (DCS) ``tree'' diagram, and may
also result from a mixing process ($D^0 \leftrightarrow \overline{D}^0$),
followed by a Cabibbo-favored (CF) decay ($\overline{D}^0 \rightarrow K^+\pi^-$).  
The DCS decay rate depends on Cabibbo-Kobayashi-Maskawa
quark-mixing matrix elements and on the magnitude of SU(3) flavor symmetry
violation~\cite{ref:Gronau-Rosner}.  Mixing may occur through two
distinct types of second-order weak processes. In the first, the $D^0$
evolves into a virtual (``long-range'') intermediate state such as
$\pi^+\pi^-$, which subsequently evolves to a $\overline{D}^0$.  The
magnitude of the amplitude for long-range mixing 
has been estimated using strong interaction models
~\cite{ref:FGLNP-2004}, but has not been determined using 
a QCD calculation from first principles.  The second type of
second-order weak process is short-range \cite{ref:Golowich_2005},
with either a ``box'' or ``penguin'' topology.  Short-range mixing is
negligible in the standard model. However, exotic weakly interacting
particles could enhance the short-range mixing and provide a signature
of new physics~\cite{beauty_petrov}.

The ratio $R$ of $D^0 \to K^+\pi^-$ to $D^0 \to K^-\pi^+$ decay rates can be
approximated 
as a simple quadratic function of $t/\tau$, where $t$ is the proper 
decay time and $\tau$ is the mean $D^0$ lifetime.  This form is valid assuming CP
conservation and small values for the parameters $x = \Delta M / \Gamma$
and $y = \Delta \Gamma / 2 \Gamma$, where $\Delta M$ is the mass
difference between the $D^0$ meson weak eigenstates, $\Delta \Gamma$
is the decay width difference, and $\Gamma$
is the average decay width of the eigenstates.
Under the assumptions stated above,
\begin{eqnarray}
R(t/\tau) = R_D + \sqrt{R_D} {y}^{\prime} \,(t/\tau) + { x^{\prime 2} + 
         y^{\prime 2} \over 4} \, (t/\tau)^2 ,
\label{eqn:R(t)}
\end{eqnarray}
where $R_D$ is the squared modulus of the ratio of DCS to CF amplitudes. 
The parameters $x^{\prime}$ and $y^{\prime}$ are linear combinations of $x$ and
$y$ according to the relations
$$
x^{\prime} = x ~{\rm cos}~\delta + y ~{\rm sin} ~\delta ~~~{\rm and}~~~ 
y^{\prime} = - x ~{\rm sin} ~\delta + y ~{\rm cos} ~\delta ,
$$
where $\delta$ is the strong interaction phase difference between the
DCS and CF amplitudes.  In the absence of mixing, $x' = y' = 0$ and $R(t/\tau) = R_D$.

We use a signal of $12.7 \times 10^3$ $D^0\rightarrow K^+\pi^-$ decays with proper decay times between 0.75 and
10 mean $D^0$ lifetimes. The data sample was recorded with the \cdf\ detector 
and corresponds to an integrated luminosity of 1.5 fb$^{-1}$. 
We search for $D^0 - \overline{D}^0$ mixing and measure the mixing parameters to be $R_D = (3.04 \pm
0.55) \times 10^{-3}$, $y' = (8.5 \pm 7.6) \times 10^{-3}$, and $x'^2 =
(-0.12 \pm 0.35) \times 10^{-3}$~\cite{Aaltonen:2007uc}.
Bayesian probability contours in the $x'^2-y'$ plane shows
that the data are inconsistent with the no-mixing hypothesis with
a probability equivalent to 3.8$\sigma$~\cite{Aaltonen:2007uc}. 
This result is in agreement with the current B-factories 
measurements~\cite{beauty_neri},
and it has a comparable sensitivity.
 
Using the current data sample, corresponding to 4.4~fb$^{-1}$ of integrated luminosity, 
we reconstruct about $24 \times 10^3$ $D^0\rightarrow K^+\pi^-$ decays. 
This is approximately a factor 2 better in statistics with respect to the published results\cite{Aaltonen:2007uc}. 
Assuming no analysis improvements and assuming as central values those published in \cite{Aaltonen:2007uc},   
we expects to reach 5$\sigma$ level of significance and to provide a precise measurement of mixing parameters $x'^{2}$
and $y'$ at the next iteration  of this analysis.
CDF is currently taking data with the goal of integrating 8~fb$^{-1}$ by the end of 2010. If the extension of Run~II
will be approved, CDF will collect on tape about 10~fb$^{-1}$, by the end of 2011, of good data for physics analysis. 
This scenario is very promising considering that CDF has already the world's largest data sample, and 
it is taking new charm data with an unprecedent rate. 


\begin{thebibliography}{99}
\bibitem{qconj}
 Throughout this paper inclusion of charge conjugate reactions are implied.
\bibitem{isodef} $\iso=| \ptmm | /(\sum_{i}p^{i}_{T} + | \ptmm | )$; the
 sum is over all tracks with $\sqrt{\Delta\eta^2+\Delta\phi^2} \le 1$;
 $\Delta\phi$ and $\Delta\eta$ are the azimuthal angle and
 pseudorapidity of track $i$ with respect to \pmm.
\end{thebibliography}
\end{document}